\documentclass[reprint,amsmath,amssymb,aps]{revtex4-2}
\usepackage{graphicx}
\usepackage{dcolumn}
\usepackage{bm}
\usepackage{amsmath}
\usepackage{caption}
\newenvironment{spmatrix}[1]
{\def\mysubscript{#1}\mathop\bgroup\begin{pmatrix}}
{\end{pmatrix}\egroup_{\textstyle\mathstrut\mysubscript}}
 
\begin{document}

\title{Flattening Surface Based On Using Contour Estimating Subdivision Surface}

\author{Yuhan Xu}
    \email{yuhanx5@uw.edu}
\author{Renqing Luo}
    \email{renqing@uw.edu}
\affiliation{Department of Applied Mathematics, University of Washington, Seattle, Washington 98195\\
Department of Physics, University of Washington, Seattle, Washington 98195}

\date{\today}

\begin{abstract}

\textbf{Abstract}: In the process of projecting the surface of a three-dimensional object onto a two-dimensional surface, due to the perspective distortion, the image on the surface of the object will have different degrees of distortion according to the level of the surface curvature. This paper presents an imprecise method for flattening this type of distortion on the surface of a regularly curved body. The main idea of this method is to roughly estimate the gridded surface subdivision that can be used to describe the surface of the three-dimensional object through the contour curve of the two-dimensional image of the object. Then, take each grid block with different sizes and shapes inversely transformed into a rectangle with exactly the same shape and size. Finally, each of the same rectangles is splicing and recombining in turn to obtain a roughly flat rectangle. This paper will introduce and show the specific process and results of using this method to solve the problem of bending page flattening, then demonstrate the feasibility and limitations of this method.
\\

\textbf{Key words: distorted image correction, document image processing, document image flattening, contour estimation for the surface subdivision}
\end{abstract}
\maketitle
\section{Introduction: Motivation and Problem Statement}
\begin{figure}
    \centering
    \includegraphics[width=0.4\textwidth]{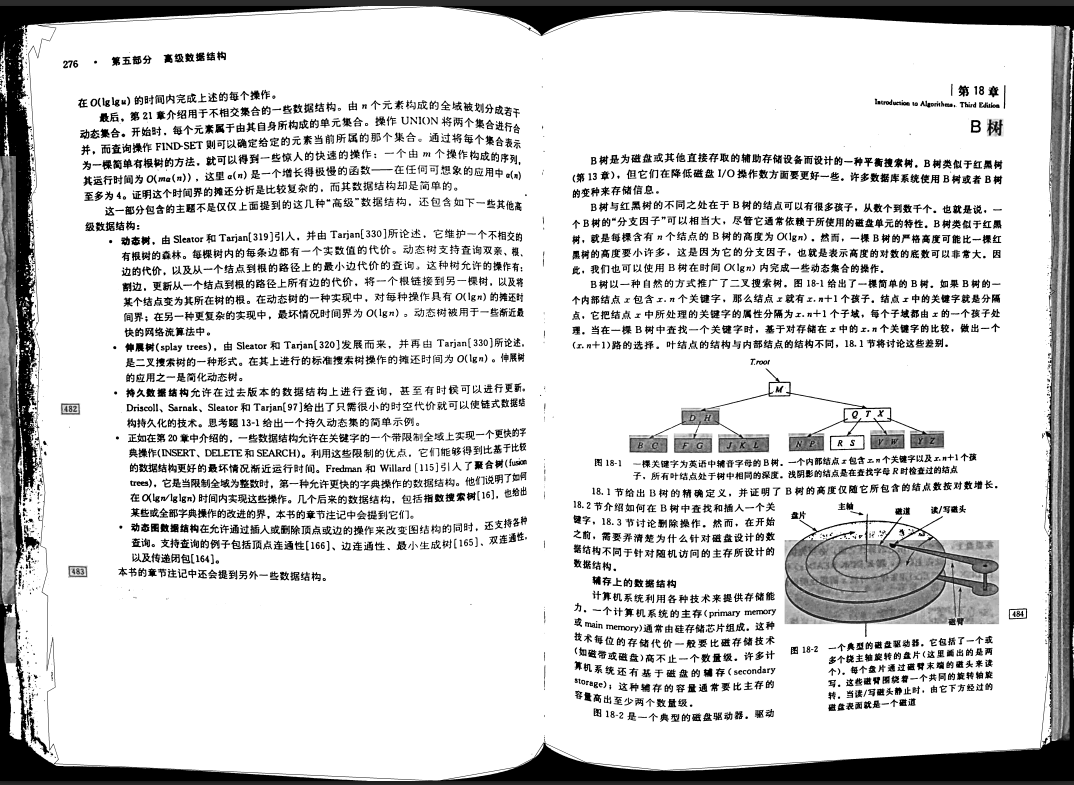}
    \caption{Preprocessed picture of a bending page}
    \label{fig:input}
\end{figure}

The method implemented in this paper was originally designed to solve the problem of image distortion caused by camera scanning curved pages. During the online class, we need to upload photos of our homework on the notebook, but the pages are bent by the stress, so the content on the page will be distorted in the photos. The closer the page is to the spine of the book in the photo, the more the text deforms. As shown in Fig.\ref{fig:input}, the deformation of the text in the image on the horizontal axis of the page will be stretched in the raised part and contracted in the concave spine part. The text will also be distorted downward or upward in the vertical axis according to the concave and convex surface. Such deformation is continuous and regulation, it is ruled by the perspective effect \cite {doi:10.1142/9789814343138_0013}. Although such deformation would not change the structure of the text, it would still greatly degrade the quality of scanned documents in electronic archives, which would make for an unpleasant reading experience. Therefore, we began to try to explore a method to solve the flattening of camera shooting curved pages at any position and any angle within a reasonable range. After finding a method to solve the problem of page flattening, we can further consider that the method can be simply extended to the problem of surface flattening of any three-dimensional objects with monotonous changes.

Before solving the problem, we took a closer look at how perspective causes distortions. Perspective simply means that the more distant an object is from the lens, the smaller the size of the image, and vice versa. The reasons for the distortion of the book page in the camera image can also be approximately explained by the perspective effect. The perspective distortion of the book can be divided into global and local. 

Global perspective distortion refers to the deformation of the overall outline of a book. This is because the lens is not perpendicular to the center of the book where the diagonal lines meet. This means that if the lens is directly above the center of the book, the four corners of the entire open book outline should be connected in a rectangle. However, if the camera is at other angles and positions, the outline of the book will be roughly irregular quadrilateral. 

\begin{figure}
    \centering
    \includegraphics[width=0.3\textwidth]{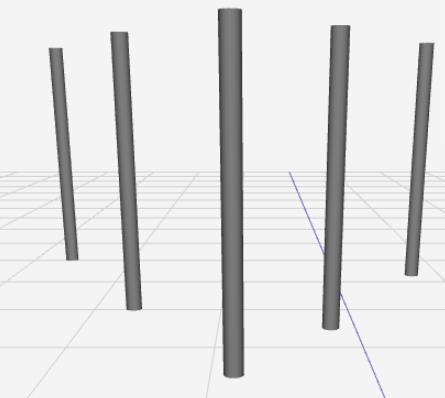}
    \caption{Analogy of contours being curved by perspective distortion}
    \label{fig:per}
\end{figure}

Local perspective distortion refers to the distortion of the local content of the text caused by the difference in the height of the page bending. Since the page itself is a flat rectangular sheet of paper, each vertical line (the long side) on the page should be the same length. But because the curvature of the horizontal axis of the page changes the distance of each long edge from the lens, the long edge near the spine is shrunk and the long edge near the bulge is stretched. As shown in Fig.\ref{fig:per}, the grey cylinders are similar to the long edges of the pages. These cylinders themselves are exactly the same, but their different distances from the lens result in different imaging sizes, and the deformation of each cylinder is the local perspective distortion.

Global perspective distortion can be solved by mapping the quadrilateral outline of the entire book to a rectangle of specified size through perspective transformation. However, the local perspective distortion is more troublesome, which is the central problem to be solved in this paper. The length change of the long side of a book caused by perspective projection is directly reflected in the bending of the outline. Therefore, we can take the shape of the outline as the entry point to find a solution to the problem of smoothing the surface of the curved text. It is not feasible to subdivide the page into strips of long edges and then stretch or shrink each strip linearly to a specified length. This is because although the bending height on the same strip is the same, the Angle of view from the top to the bottom of the lens changes, resulting in the actual distance from the local page to the lens being different.

The main idea of solving the problem is to determine the perspective principle as the basic principle, which can explain the distortion phenomenon of text image. Then, the distortion problem caused by the complex page bending height difference is subdivided into subproblems that can be solved by the basic principles. Finally, all the basic subproblems are spliced back together, so that the complex problem can be solved approximately. 

We know the most basic solution to quadrilateral perspective distortion, so we can subdivide the surface into discrete quadrilateral blocks, and then invert perspective into equal small rectangles for each block, and finally splicing all the small rectangles into one big block in turn. This is a very simple and familiar classical idea, but there are some technical problems with the way this article solves the problem. 

For example, how to specifically determine the subdivision rules of the surface through the shape of the contour, so that the subdivision surface can be restored to the real three-dimensional surface as far as possible. This is also the focus of the method of this paper and related works.

\begin{figure}
    \centering
    \includegraphics[width=0.3\textwidth]{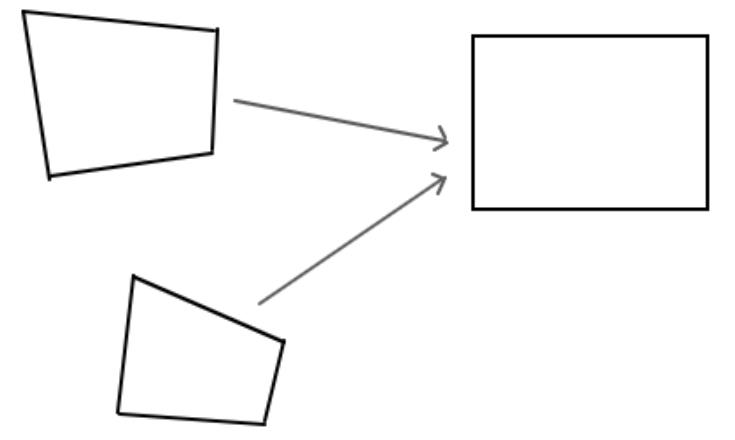}
    \caption{Quadrilateral inverse perspective to rectangle}
    \label{fig:trans}
\end{figure}

\subsection{Related Works}

There are various solutions to flatten the two-dimensional projection image of three-dimensional object surface, and the different solutions are mainly determined by the specific application scene. There are two main conditions for the application of the method in this paper: 1. Single imaging images with single lens at any Angle and distance. 2. The leveling algorithm meets the requirements of simple implementation, timeliness and lightweight. 

A method to restore surface shape by texture flow field, parallelism, and equal spacing of text content on the page\cite{1467462}, satisfies the first condition we need, but its algorithm implementation will be complicated. Another document, \cite{1227630}, can correct the distortion more accurately by using perspective geometry combined with the cylindrical modeling of page bending. However, this method needs to provide the distance and Angle of the camera for the calculation of perspective geometry. Therefore, it is suitable for high camera with fixed camera position, rather than mobile phone shooting at arbitrary position. There is also a method to reassemble multiple key frames by shooting continuous video to achieve the effect of correcting the distortion of the document \cite{7761529}. 

There are open source projects that use minimal optimization algorithms to correct the distortion \cite{a2016_page}, but this method requires a lot of computing resources and time. There are also some methods assisted by additional equipment, such as laser grid 3D reconstruction \cite{6909892} or stereo camera \cite{1334171}. The solution presented in this article may not be as effective as the above method, but it is very simple to use, only requires a camera phone, and the implementation of the algorithm is very simple.

\section{Solution to Problem}
This paper has introduced the basic idea of problem solving before, here we will provide the detailed steps of the method and the specific technical details of the problem solving. The steps of this method can be divided into four parts: 1. Contour preprocessing; 2. Contour function fitting; 3. Gridded surface, 4. Recombination after perspective transformation.

In contour preprocessing, we first extract the boundary contour of two-dimensional surface image, and then divide the image into several smooth surfaces according to the contour. Finally, each smooth surface is divided into four edges. For contour function fitting, we first sampled coordinate points for each boundary contour according to the specified precision, then adopted appropriate functions and fitted within the boundary range, and finally evolved continuously from the fitting function of the upper/left boundary line to the fitting function of the lower/right boundary line. For gridded surfaces, the sampling spacing of the function family is determined according to the curvature of the boundary and its positive and negative values. Then, the quadrilateral mesh points of subdivided small surfaces can be obtained from the intersection points of the upper, lower, left and right function families. After the final perspective transformation, we first transform each grid into a small rectangle with the same size and shape, and then spliced each small rectangle after the perspective transformation into a complete large rectangle in turn, so that we can get a flattened image with distorted content.

\subsection{Contour Preprocessing}
\begin{figure}
    \centering
    \includegraphics[width=0.2\textwidth]{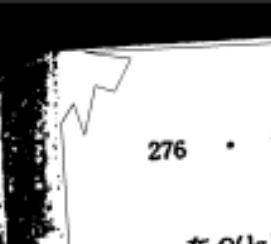}
    \caption{Outline of abnormal contour. Black line is contour.}
    \label{fig:bad}
\end{figure}
Since we need to get the curvature of the surface from the outline of the page, we first need to extract the outline of the page in the photo. This paper adopts the contour extraction algorithm \cite{SUZUKI198532}, which is implemented in OpenCV \cite{opencv_library}. This algorithm requires that the pages and background should be clear and have obvious black-and-white contrast when we take photos. Since the background needs to have a sharp contrast between black and white and the page, in practice we also need to binarize the Fig.\ref{fig:input} first. For the book document, in the actual processing, we first extract the quadrilateral approximate outline of the entire book, and then transform the quadrilateral into a rectangle, so as to complete the correction of the global perspective.

After completing the extraction of the outline, we can then see that there is a continuous but undifferentiated point in the middle of the outline of the book, that is, the upper and lower boundary at the spine. Such profiles will greatly interfere with our subsequent contour fitting, and we cannot calculate the curvature on a function with underivable points. So next we need to divide the contour into two smooth surfaces, the left and right pages. The contour extracted in the computer is not smooth originally, but the approximation of the curved contour by discrete points. Therefore, we can easily obtain the slope of the line segment connected by every two points on the approximate curve, and then calculate the change of the slope from the current line segment to the next line segment in turn, and set a threshold. If the threshold is exceeded, it can be found that the intersection point of two line segments is an underivable point.

In actual processing, we need to narrow the scope of finding underivable points as much as possible, because as shown in Figure \ref{fig:bad}, contour extraction depends on the quality of the picture, there will be unreasonable multiple underivable points. The four corners of a book page are not derivative points, but we do not use this method to obtain, but we have roughly extracted the quadrilateral outline of the book itself is made up of these four corners. Finally, according to these four points plus the two undifferentiated points in the middle, we can divide the page into two smooth surfaces, and both the left and right smooth surfaces can be divided into four smooth boundary lines. And then we can fit these boundaries.
\subsection{Contour Function Fitting}
The fitting of contour function is to let the boundary curve evolve from top to bottom and from left to right and calculate the curvature of the contour at each position. The curve evolution is to solve the problem of local characters bending up and down the vertical axis, and the curvature calculation is to solve the problem of the contraction or stretching of the horizontal axis. The boundary we obtained before is actually composed of discrete points, and the number of points depends on the sampling accuracy of contour extraction. Therefore, we need to use these discrete points to fit a smooth function. Here, we choose to use a polynomial function of order $n$ to fit the boundary contour.
\begin{equation}
P(x) = c_n x^n + c_{n-1} x^{n-1}+\cdots+c_1 x+c_0
\end{equation}
This is because polynomial functions can be fitted very quickly, arbitrary smooth curves can be fitted in a finite small range, and the calculation of curvature is extremely simple. After obtaining the polynomial expression of the boundary line, we can calculate the family of evolutionary polynomial functions to simulate that the local distortion becomes more obvious the further away from the lens under perspective effect. This evolution is indeed not linear. For example, turning a laser beam just a little at a large Angle will greatly vary the length of the sweep and the distance from the end point to the laser. In other words, if the lens rotates $\theta$ correctly, the ratio of the lens passing distance to the distance from the end point to the lens should be $\sin \theta$, where $\theta$ is the Angle between the two lines. But in the case of the pages we can use a small Angle approximation, so $\sin \theta \approx \theta$. Because accurate calculation requires knowing the vertical distance from the lens to the page, we choose an approximately uniform evolutionary boundary fitting function.
\subsection{Gridded Surface}

Gridded surface is a method used to describe the projection of the surface of a three-dimensional object onto a two-dimensional surface, in which each subdivision block can be regarded as the result of perspective transformation on the position of a rectangle evenly divided after the original flat surface. This paper will roughly estimate the rules of subdivision surfaces by the page edge contour. Among them, the evolution of fitting function family according to edge contour and the calculation of curvature are the most critical steps in the whole method, because this step determines the rules of surface subdivision, which directly determines the final flattening effect and quality. The gridded surface is formed by a series of evolved polynomial functions crossing each other on the horizontal and vertical axes. The sampling interval of the evolved functions is determined by the curvature and concavity of the boundary functions.

\begin{figure}
  \begin{minipage}[t]{0.5\linewidth}
    \centering
    \includegraphics[scale=0.17]{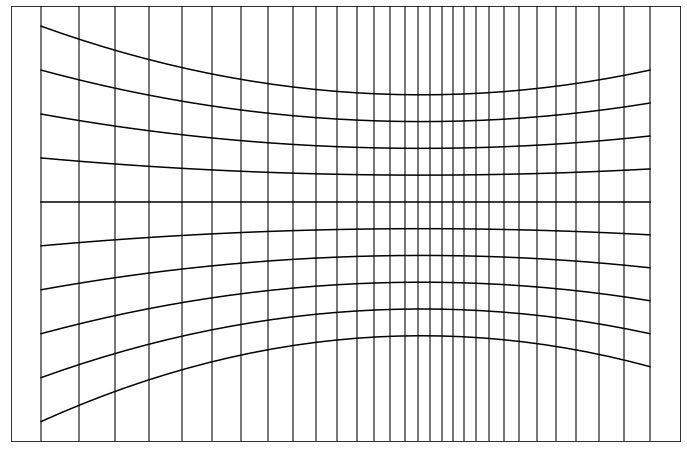}
    \caption{Sunken Inward}
    \label{fig:ao}
  \end{minipage}%
  \begin{minipage}[t]{0.5\linewidth}
    \centering
    \includegraphics[scale=0.17]{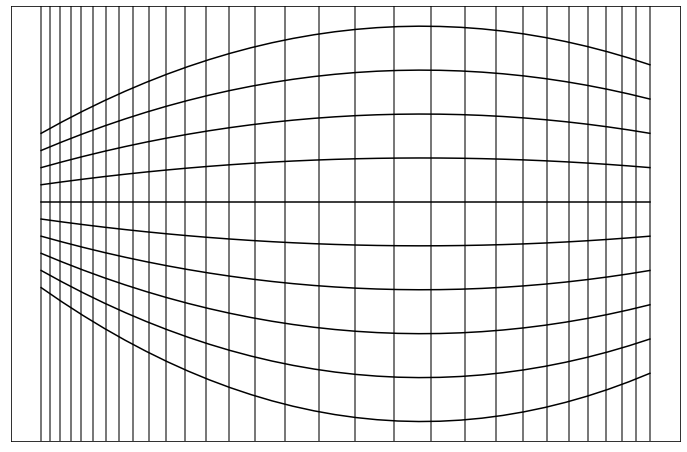}
    \caption{Protruding Outward}
    \label{fig:tu}
  \end{minipage}
\end{figure}

\subsubsection{Evolution of Polynomial Functions}

We have discussed why we can use uniform polynomial function evolution, and here we will explain in detail how we can evolve polynomial functions. A polynomial function can evolve evenly into another polynomial function by adjusting its coefficients, as long as the highest order of the two polynomial functions are guaranteed to be equal. So we can make each coefficient of the polynomial uniformly change to the corresponding coefficient of the other polynomial according to the specified step size. To be precise, the polynomial function $P_0(x)$ uniformly evolves to the step size of $P_n(x)$ as,
\begin{equation}
d = \frac{P_n(x)-P_0(x)}{n+1}
\end{equation}
Where $n+1$ is the number of polynomial evolution (steps). According to this step size, the stepping rule of uniform evolution of polynomial function is,
\begin{equation}
P_{i+1}(x) = P_{i}(x) + d
\end{equation}
here $i$ is a positive integer index ranging from $0$ to $n$. Although the evolution of the polynomial is uniform, we need to consider the non-uniform sampling interval caused by curvature, so we can set a larger number of evolution steps to make the step size small, so that the accuracy of the following interval sampling can be improved.

\subsubsection{Calculation of Concavity and Curvature}

Curvature is calculated to solve the shrinkage or stretching of the page in the lateral direction, and it requires the first and second derivatives of the function, so the polynomial function we need to fit is at least a second order polynomial. In practice, however, we should take higher order polynomials to deal with more complex bending shapes. Here we do not care about the intermediate family of function evolution, the calculation of curvature only works on the fitting function of the boundary contour. Generally speaking, the boundary curve of a page has only one extreme point, so we can judge the convexity of the function by the second derivative. However, if the page is repeatedly bent with multiple extreme points, then we need to further divide the curve into concave-convex functions based on the inflection points. The inflection point can not be determined by discrete sampling points, but needs to solve the second derivative of the analytic expression of the fitting polynomial function, that is, to solve the value of $x$ for the equation $P''(x) = 0$. Next, for each boundary polynomial curve $P(x)$ divided according to the inflection point, we judge the positive and negative of its second derivative in the partition range to determine the convexity of the function. The curvature $\kappa$ of the boundary function of the horizontal axis at the point $x$ is,
\begin{equation}
\kappa(x) = \frac{P''(x)}{(1+P'(x))^{3/2}}
\end{equation}
Curvature describes the degree to which a curve is bent. The higher the page is raised, the more the text is stretched, the more curved the outline appears, and the greater the absolute value of the curvature here. Therefore, the sampling interval of the evolution function should be stretched where the absolute value of curvature is large, so that the perspective of the inverse transformation can shrink the stretched text to its original size as much as possible.
\begin{figure}
    \centering
    \includegraphics[width=0.35\textwidth]{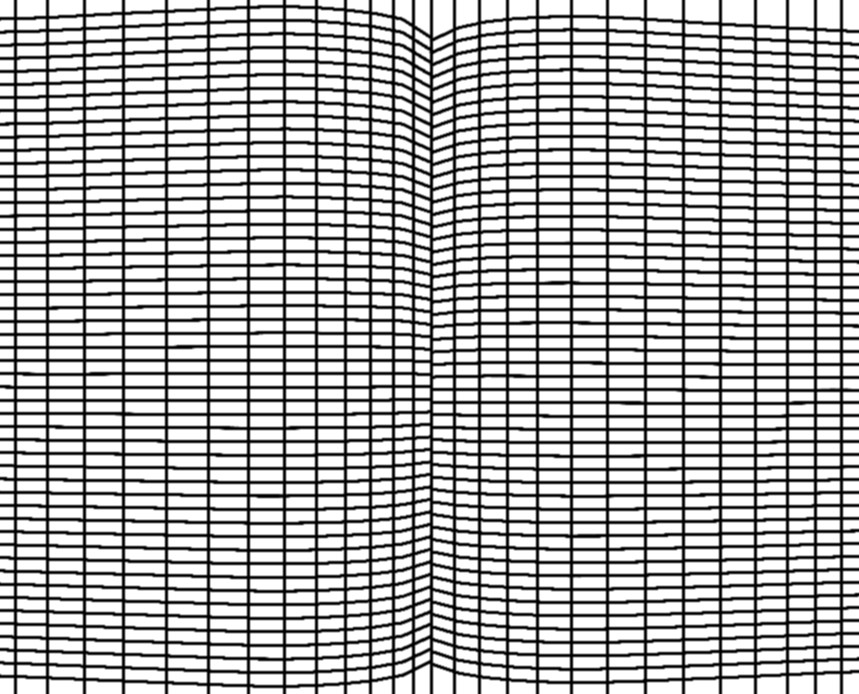}
    \caption{Approximate curved subdivision of a page}
    \label{fig:gri}
\end{figure}
The degree of curvature of the function is the same no matter it is concave or convex. For example, the absolute value of curvature of $f(x) = x^2$ and $f(x) = -x^2$ are the same at each point, but the difference of convexity on both sides of the boundary function should affect the value of curvature. As shown in Fig.\ref{fig:ao}, although the absolute values of curvature of the upper and lower boundary functions are the same, the sampling interval of function evolution is reduced in places with large curvature instead of increasing the sampling interval as shown in Fig.\ref{fig:tu}. Therefore, it is not difficult to see from Fig.\ref{fig:ao} that when the convex function evolves to concave function, the sampling interval will shrink in the place where the curvature is larger. From Fig.\ref{fig:tu}, it can be seen that when the concave function evolves to convex function, the sampling interval will expand with the increase of curvature. Furthermore, the less the boundary function changes from top to bottom, the more evenly spaced the samples can be imagined. That is to say, in fact, in a concrete implementation, we can't just consider the curvature of the boundary function on one side, but we need to calculate the average of the difference between the curvatures on both sides,
\begin{equation}
k = \frac{\kappa_f-\kappa_i}{2}
\end{equation}
Finally, the scaling factor of sampling interval is determined to be
\begin{equation}
\gamma = |k|^{-\frac{k}{|k|}}
\end{equation}
Where $\frac{k}{|k|}$ is used to judge the $k$ symbol of positive and negative, Fig \ref{fig:ao} the curvature of the upper boundary function $\kappa_i$ and lower boundary curvature $\kappa_f$ gives $\kappa_i = - \kappa_f $, the average curvature difference is $k=\kappa_f$, and the scaling coefficient is $\gamma=\kappa_f^{-1}$ when the curvature of the lower boundary function is positive. Hence, the Fig.\ref{fig:ao} shows a denser grid distribution where the absolute value of curvature is larger. Similarly, the scaling factor calculated in Fig.\ref{fig:tu} can be obtained as $\gamma=\kappa_f$, resulting in a more sparse grid distribution where the absolute value of curvature is larger.

Finally, with the evolution of the fitting function and the calculation of curvature, we can deal with the concrete meshing. Let $M$ be the sampling accuracy (number) of the evolution function family on the horizontal axis, and $N$ be the sampling accuracy of the vertical axis. When calculating the curvature on the boundary curve of the horizontal axis, it needs to be divided into $N$ uniform sampling points $x_i$ according to the sampling accuracy of the vertical axis. Then, for each sampling point, we obtain the scaling coefficient $\gamma_{i}$. Also have all the scaling coefficients on the horizontal axis of the number of listed as $\{\gamma_{i}\}_{i=0}^{N-1}$, scaling coefficient on the vertical axis number listed as $\{\gamma_{j}\}_{j=0}^{M-1}$. Next, we need to correspond these scaling coefficients to the scale of the entire horizontal axis, so we need to normalize the scaling coefficient series.
\begin{equation}
\{\Gamma_i\} = \frac{\{\gamma_{i}\}_{i=1}^{N-1}}{\sum_{i=1}^{N-1} \gamma_i}
\end{equation}
Similarly, $\{\Gamma_j\}$ can be obtained after the normalization of the vertical axis. Notice that we exclude $i=0$ from the normalization process, because when we calculate the horizontal sampling interval point $X$,
\begin{equation}
X_{i+1} = X_{i} + \Gamma_i(x_{N-1}-x_0), \ X_0 = x_0
\end{equation}
Here $i=0$ is the starting point and does not participate in the calculation of the scaling factor. Finally, we can calculate each longitudinal evolution function of the sequence $\{X_i\}_{i=0}^{N-1}$ according to the sequence of interval points on the horizontal axis. Similarly, we can calculate all the corresponding horizontal evolution functions of the sequence $\{Y_j\}_{j=0}^{M-1}$ on the vertical axis. In this way, the crisscross evolution function family completes the division of subdivision surface. Fig.\ref{fig:gri} is the subdivision estimation of Fig.\ref{fig:input}. Although the partition surface is an imprecise estimate based on the contour and its curvature, good results can be obtained after completing the last step of the following.
\begin{figure}
    \centering
    \includegraphics[width=0.4\textwidth]{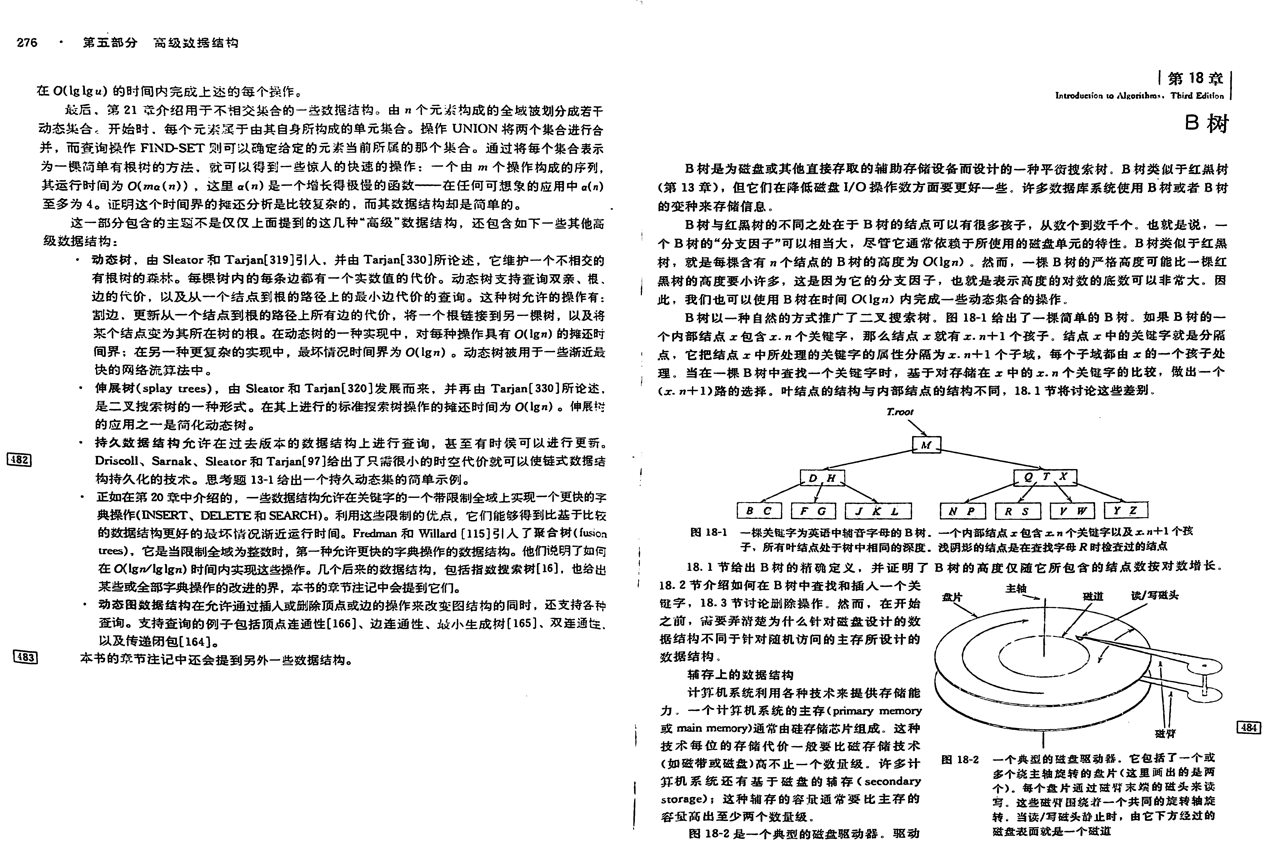}
    \caption{The result of the flattening of a book page}
    \label{fig:out}
\end{figure}
\subsection{Reconfiguration after Inverse Perspective}
The idea and principle of recombination is the most simple and direct in the whole method, and it is also a step with correct and single implementation. All we need to do here is to splice each quadrilateral block in the previous surface subdivision result back to a complete image after the perspective inversion. However, this part is also the most cumbersome part in the implementation of the whole method, the most errors caused by the index, the largest amount of code, the longest running time, and the most need for parallel optimization. First of all, we only give the sampling and spacing of the family of vertical and horizontal functions in the grid division before, but the computer does not directly know each quadrilateral block, so it cannot process it. Therefore, we first need to compute the intersection points of all the families of vertical and horizontal functions from left to right and from top to bottom, and store each point $p_k=(x_k, y_k)$ in a list $\{p_k\}_{k=0}^{MN-1}$. Then, the quadrilateral block is divided according to the index of intersection points and the number of samples. Based on the lower-left corner of the quadrilateral, we formed a quadrilateral block with four points left, up and up. Since there was no next point right at the lower-left corner of the right-most boundary, and no next point upward at the top-most boundary, we did not extract quadrilateral blocks for the upper and right boundary points. To be precise, the point $p_k$ has the corresponding quadrilateral block of
\begin{align}
B_k = \{p_k, p_{k+1}, p_{k+M}, p_{k+M+1}\},\\
\mod(k)\neq M-1 \ and \ k<M(N-1).
\end{align}
Next, after we have obtained all the quadrilateral block lattice, we can map the quadrilateral to the congruent rectangle in turn by the perspective inverse transformation. In this paper, the processing of inverse perspective transformation is done by the built-in function in OpenCV \cite{opencv_library}. The principle is to solve the $3\times 3$ perspective transformation matrix $M$ in the following equation, where $t_i=x_ih_7+y_ih_8+h_9$ is the scaling coefficient. $(x_i y_i), \ i = 0,1,2,3 $ for input quadrilateral four corner point, $(x'_i, y'_i)$ for the output quadrilateral points at the edges.
\begin{equation}
    \begin{bmatrix} t_i x'_i \\ t_i y'_i \\ t_i \end{bmatrix} = \begin{spmatrix}{M} h_1 & h_2 & h_3 \\ h_4 & h_5 & h_6 \\ h_7 & h_8 & h_9 \end{spmatrix} \cdot \begin{bmatrix} x_i \\ y_i \\ 1 \end{bmatrix}
\end{equation}
Next, according to the perspective transformation matrix $M$, we can perspective the image content (all pixels) on the corresponding quadrilateral block to the target congruent rectangle\cite{arcangelodistante_2021_handbook}. Finally, each rectangle obtained through the inverse perspective transformation is spliced and reassembled into a large rectangular plane in sequence, and the whole leveling is completed.

Almost all operations in this step are repeated and independent. The calculation on each function is the same and repeated when extracting a lattice of intersection of evolution functions, in addition to obtaining each block with a lattice and calculating the inverse perspective transformation for each block. The repeated and independent computations can be processed simultaneously, thus greatly improving the processing efficiency of the computer and reducing the time cost of the algorithm.
\section{Results and Analysis}
\begin{figure}
    \centering
    \includegraphics[width=0.4\textwidth]{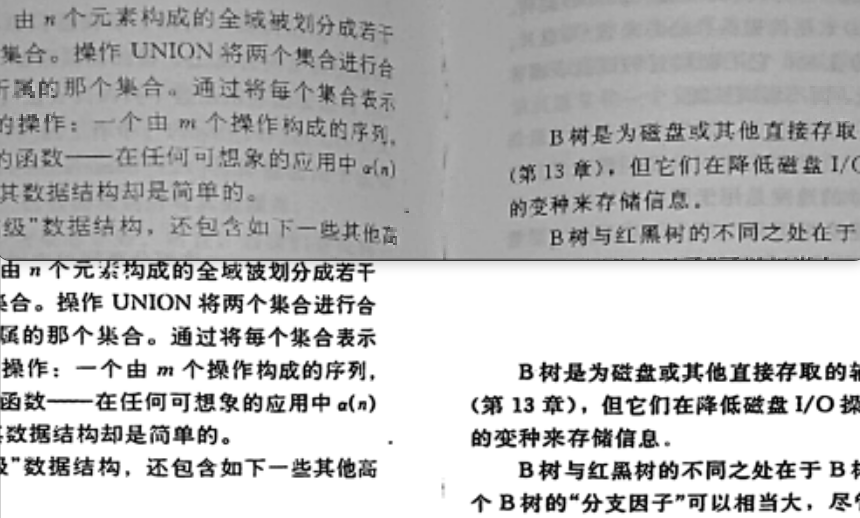}
    \caption{Comparison of partial page bending and flattening before and after treatment}
    \label{fig:comp}
\end{figure}

As shown in Fig.\ref{fig:out}, this is the result obtained after leveling the bent pages in Fig.\ref{fig:input} by the above method. Fig.\ref{fig:comp} is the local comparison diagram at the spine of the book. Here we can see that after the inverse perspective transformation of each grid, we successfully flattened the original distorted sentences into straight and parallel sentences, which significantly improved the vertical bending and horizontal scaling of the text. However, as this article solves the problem of shrinkage of the vertical axis by taking an imprecise estimate, we can see that the text is still slightly inconsistent in size in different places. In addition, our original code implementation assumed that the vertical axis of the page was vertical, but in fact there was a slight slant on the left edge of the Fig.\ref{fig:input}, which resulted in some slight distortion visible at the bottom of the left page near the spine.
In addition, the character recognition rate (CRR) in the distorted part of the document in Fig.\ref{fig:comp} is $81.73\%$, and the CRR after flattening is $96.28\%$. However, for OCR \cite{Zhang_2020_CVPR} based on deep learning text detection, this method will not provide a significant improvement, because distorted documents also have a very high character recognition rate after text detection. While this approach does not improve modern OCR significantly, it does improve the reading experience by smoothing out non-text content such as handwritten content, formulas, and images on curved pages.

\section{summary}
In recent decades, with the development of digitization and information technology, document digitization and document image processing have more and more important practical and research value. The correction of distorted documents is also changing rapidly, and a variety of solutions should be available to solve the problem in different scenarios. Based on the outline of the page, this paper solves the flattening problem of the curved page in the single shot imaging. The distortion correction method proposed in this paper can not only be applied to the distortion correction of text documents, but also to the correction of distorted images on arbitrary monotonic deformation surfaces. At the same time, this method has very low requirements on hardware equipment, and only needs to shoot a single photo with a single lens. Moreover, the algorithm complexity of this method is also relatively low, and the implementation of the algorithm is also very simple. However, this method is an inexact correction because it uses the curvature of the contour to estimate the segmentation of the finite mesh. Although the flattening effect of this method has a good performance in experimental tests, this estimation method does not have strict theoretical support. The main limitation of this method is that the outline of the document needs to be very clear, and it cannot handle the flattening of the documents with folds and irregular distortion. Its main advantage is that it can solve the distortion problem of most daily text photography scanning by using a software with low memory consumption and low computational power demand.
\bibliography{apssamp}
\end{document}